\documentclass[aps,pre,showpacs]{revtex4}
\usepackage{amssymb}

\usepackage{amsmath}
\usepackage{graphicx}

\newcommand{\be}{\begin{equation}}
\newcommand{\ee}{\end{equation}}

\begin{document}

\title{Fluctuating Filaments Under Tension - From Flexible Chains to Rigid
Rods }
\author{David A. Kessler and Yitzhak Rabin}
\affiliation{Department of Physics, Bar--Ilan University, Ramat-Gan 52900, Israel}

\begin{abstract}
We develop a novel biased Monte-Carlo simulation technique to measure the
force-extension curves and the distribution function of the extension of
fluctuating filaments stretched by external force. The method is applicable
for arbitrary ratio of the persistence length to the contour length and for
arbitrary forces. The simulation results agree with analytic expressions for
the force-extension curves and for the renormalized length-scale-dependent
elastic moduli, derived in the rigid rod and in the strong force limits. We
find that orientational fluctuations and wall effects produce non-Gaussian
distributions for nearly rigid filaments in the small to intermediate force
regime. We compare our results to the predictions of previous investigators
and propose new experiments on nearly rigid rods such as actin filaments.
\end{abstract}

\date{\today }
\maketitle

\section{Introduction}

Recent experimental studies of the mechanical properties of biological
materials, ranging from individual DNA molecules\cite{bustamante} to single
actin filaments and their networks\cite{albert,fred,sackmann}, led to
revival of interest in the worm-like chain model of semiflexible
biopolymers. In particular, the force-extension curves predicted by this
model in the long chain limit (i.e., when the length of the chain is much
larger than the persistence length) were shown to agree quantitatively with
experimental results on dsDNA at large elongations\cite{ms1}. However, many
interesting questions concerning worm-like chains under tension are still
open and there are delicate issues concerning even the linear response of
the worm-like chain to stretching force\cite{frey2}. Also, while the
equilibrium distribution function of the extension was calculated for the
worm-like chain\cite{frey1,sinha,dhar} and fluctuating ribbon\cite{jcp-kr}
models, only approximate (using the Gaussian approximation) calculations of
the distribution function for stretched chains were reported to date\cite%
{winkler}. Such distributions have been measured in experiments on stretched
DNA molecules\cite{thesis} and provide a much more stringent test of the
theory than the force-extension curves. Finally, even though all mechanical
experiments involve biopolymers attached to surfaces, none of the present
theories incorporates wall effects on force-extension curves and on the
corresponding distributions. While such effects are negligible for nearly
fully stretched chains, they are expected to be important in the weak force
regime.

In previous work\cite{Kats,jcp-kr}, we have discussed how to generate via
Monte-Carlo equilibrium ensembles of fluctuating ribbons with spontaneous
curvature and/or twist. This method was based on using the intrinsic
geometry of the filament, in which representation the energy is purely
quadratic. This enables a direct generation of independent samples
distributed according to the Gibbs equilibrium ensemble. When external
forces act on the filaments, the energy is no longer a quadratic function of
the curvatures, and the problem is now not only nonlinear but also nonlocal.
In this paper we discuss a biasing method which succeeds in efficiently
generating Monte-Carlo samples for the simplest case of an external force,
namely that of the stretched filament. This problem was studied by Marko and
Siggia, (MS), \cite{ms1,MS94} and others \cite{odijk,ha} who calculated the
force-extension curve for long worm-like chains. We will reproduce these
results as well as extend the analysis to the case of nearly stiff filaments
with length comparable to or shorter than the persistence length, and in
addition calculate the distribution of extensions for various forces.
Lastly, we will show how we can also easily incorporate steric constraints;
we will demonstrate this in the context of a restriction of the filament to
a half-space by the presence of a wall.

In sections 2 and 3 we review the general theory of fluctuating ribbons (we
define a ribbon as a filament with an asymmetric cross--section) and the
Frenet algorithm for computer simulations of such ribbons, respectively. In
section 4 we describe the biased Monte-Carlo simulations of filaments under
tension and present the resulting force-extension curves and distribution
functions. Wall effects are considered in section 5. In section 6 we discuss
our results and outline directions for future research. In Appendix A we
expand the Euler angles in terms of the principal curvatures and then use
this expansion in Appendix B in order to calculate the partition function of
a stretched filament. In part 1 of this appendix we derive an analytic
expression for the partition function and obtain a force-extension relation
for arbitrary force in the limit of large persistence length (expansion
about the rigid rod limit). In part 2 of the appendix we derive asymptotic
expressions for the above quantities in the limit of large force (for
arbitrary rigidity). Finally, in Appendix 3 we present the calculation of
the renormalization of the elastic moduli of the various Fourier modes of
the curvature.

\section{General theory}

The linear elasticity of ribbons is described by the elastic energy\cite%
{Helixpre} 
\begin{equation}
E_{el}=\dfrac{1}{2}\sum_{k=1}^{3}b_{k}\int_{0}^{L}ds\left(
\omega_{k}-\omega_{0k}\right) ^{2},   \label{energy}
\end{equation}
where $s$ is the distance measured along the contour of the ribbon, the
coefficients $b_{1}$ and $b_{2}$ are the moduli associated with bending
along the two principal axes of inertia of the cross section (they differ if
the cross section is not circular), and $b_{3}$ is the twist modulus. In
this paper we treat $\left\{ b_{i}\right\} $ as given material parameters of
the ribbon. The functions $\left\{ \omega_{k}(s)\right\} $ and $\left\{
\omega_{0k}(s)\right\} $ are the generalized curvatures in the deformed and
the stress-free states of the ribbon, respectively. These curvatures
completely determine the three dimensional conformation of the ribbon
through the generalized Frenet equations\cite{Helixpre} 
\begin{equation}
\dfrac{d\mathbf{t}_{i}(s)}{ds}=-\sum_{j,k}\varepsilon_{ijk}\omega _{j}(s)%
\mathbf{t}_{k}(s).   \label{Frenet}
\end{equation}
Here $\mathbf{t}_{3}$ is the unit tangent to the centerline and the unit
vectors $\mathbf{t}_{1}$ and $\mathbf{t}_{2}$ are oriented along the
principal axes of the cross section ($\varepsilon_{ijk}$ is the
antisymmetric unit tensor). Note that since these equations describe a pure
rotation of the triad $\left\{ \mathbf{t}_{i}(s)\right\} $ as one moves
along the contour of the ribbon, the constraint $\left| \mathbf{t}%
_{3}\right| =\left| d\mathbf{r(}s)/ds\right| =1$ is automatically satisfied
in this intrinsic coordinate description.

The physical frame of the ribbon $\left\{ \mathbf{t}_{i}(s)\right\} $ can be
related to the original Frenet description of space curves in terms of a
unit tangent (which coincides with $\mathbf{t}_{3}$), normal $\mathbf{n}$
and binormal $\mathbf{b}$, provided that in addition to the Frenet triad $%
\left\{ \mathbf{b}(s),\mathbf{n}(s),\mathbf{t}_{3}(s)\right\} $ associated
with the centerline of the ribbon, one introduces an angle $\alpha (s)$ that
describes the rotation of two of the principal axes of the cross section $%
\mathbf{t}_{1}$, $\mathbf{t}_{2}$ with respect to the binormal $\mathbf{b}$
and normal $\mathbf{t}$ (see Fig. 1 in the second of Refs. [%
\onlinecite{Helixpre} ] ), 
\begin{equation}
\mathbf{t}_{1}=\mathbf{b}\cos \alpha +\mathbf{n}\sin \alpha ,\qquad \mathbf{t%
}_{2}=-\mathbf{b}\sin \alpha +\mathbf{n}\cos \alpha .  \label{t1n}
\end{equation}%
The three dimensional configuration of the centerline is determined by the
Frenet equations 
\begin{equation}
\frac{d\mathbf{b}}{ds}=\tau \mathbf{n},\qquad \frac{d\mathbf{n}}{ds}=\kappa 
\mathbf{t}_{3}-\tau \mathbf{b},\qquad \frac{d\mathbf{t}_{3}}{ds}=-\kappa 
\mathbf{n}  \label{Fre2}
\end{equation}%
where the curvature $\kappa (s)$ and torsion $\tau (s)$ are treated as known
functions of position along the contour. The three generalized curvatures $%
\left\{ \omega _{k}(s)\right\} ,$ can be expressed in terms of these
parameters and the angle $\alpha (s),$%
\begin{equation}
\omega _{1}=\kappa \cos \alpha ,\qquad \omega _{2}=\kappa \sin \alpha
,\qquad \omega _{3}=\tau +d\alpha /ds.  \label{frames}
\end{equation}

\section{The Frenet algorithm}

\label{simulation}

Since the energy is a quadratic form in the deviations $\delta\omega_{k}$ of
the generalized curvatures from their values in the stress free state, Eq. (%
\ref{energy}) is valid as long as the characteristic length scale of the
deformation is much larger than the diameter of the ribbon\cite{Love}. When
a ribbon undergoes fluctuations in the presence of a thermal bath at
temperature $T,$ this energy determines the statistical weight of the
configuration $\left\{ \omega_{k}\right\} $. The statistical average of any
functional of the configuration $A(\left\{ \omega_{k}\right\} )$ is defined
as the functional integral\cite{Helixpre} 
\begin{equation}
\left\langle A\left( \left\{ \omega_{i}\right\} \right) \right\rangle =\frac{%
\int D\left\{ \delta\omega_{i}\right\} A\left( \left\{ \omega _{i}\right\}
\right) e^{-E_{el}\left\{ \delta\omega_{i}\right\} /kT}}{\int D\left\{
\delta\omega_{i}\right\} e^{-E_{el}\left\{ \delta\omega _{i}\right\} /kT}%
}.   \label{statav}
\end{equation}
where $k$ is the Boltzmann constant. From the form of the energy, Eq. %
\ref{energy}, it follows that the distribution of $\delta\omega_{k}$ is
Gaussian, with zero mean ($\left\langle \delta\omega_{i}(s)\right\rangle =0$%
) and second moments given by

\begin{equation}
\left\langle \delta\omega_{i}(s)\delta\omega_{j}(s^{\prime})\right\rangle =%
\dfrac{kT}{b_{i}}\delta_{ij}\delta(s-s^{\prime}).   \label{corr}
\end{equation}

In order to study the distribution functions of the various fluctuating
quantities (e.g., the extension $R$), we developed what we entitled
the Frenet algorithm,
which is based on the combination of statistical methods and differential
geometry and is described in the following. Examination of Eq. (\ref{corr})
shows that the decoupling of the ``noises'' $\left\{
\delta\omega_{k}\right\} $ in the intrinsic coordinate representation
permits an efficient numerical generation of independent samples drawn from
the exact canonical distribution. The Gaussian distribution of the $%
\delta\omega_{k}$ means that each $\delta\omega_{k}(s)$ can be directly
generated as one of a string of independent random numbers drawn from a
distribution symmetric about the origin with width $\sqrt{%
kT/(b_{k}\Delta s)}$ where $\Delta s$ is the discretization step length.
The remaining task is to construct the curve using the Frenet equations with 
$\omega_{k}(s)=\omega_{0k}(s)+\delta\omega_{k}(s)$. The Frenet equations are
best integrated by stepping the basic triad $t_{k}$ forward in $s$ through a
suitable small rotation. In this way, the orthonormality of the triad is
guaranteed to be preserved up to machine accuracy. To construct this
rotation matrix, we begin with Eqs. (\ref{Frenet}) and, defining the three
vectors $v^{x}=(t_{1}^{x},t_{2}^{x},t_{3}^{x})$, and so forth, we can write
this equation as 
\begin{equation}
\frac{\Delta v^{i}}{\Delta s}=Av^{i},   \label{Aeq}
\end{equation}
where $A$ is an antisymmetric matrix with elements $A_{ij}=\sum_{k}%
\varepsilon_{ijk}\omega_{k}$. We now discretize Eq. (\ref{Aeq}) as 
\begin{equation}
v^{i}(s+\Delta s)=Ov^{i}(s),
\end{equation}
where the orthogonal matrix $O$ is 
\begin{equation}
O=\left( 1+\frac{\Delta s}{2}A\right) \left( 1-\frac{\Delta s}{2}A\right)
^{-1}.
\end{equation}
The choice of discretization step length ($L/\Delta s$ is the number of
points) depends on the amplitude of the random noise, $\sqrt{%
kT/(b_{k}\Delta s)}.$ We found that $400$ points gave sufficient
accuracy (i.e., was limited by the statistical noise determined by the
number of independent samples) for the calculation of the distribution
function of the extension, when the smallest of the moduli satisfied $%
kTL/b_{k}^{\min }\lesssim1.$

This algorithm is optimal for studying the fluctuations of a free ribbon.
However, if the ribbon is acted upon by external forces and/or constraints,
the algorithm must be modified. This is described in the next section.

\section{Biasing the Monte-Carlo}

Consider a filament with a circular cross--section ($b_{1}=b_{2}$) which is
characterized by a bending persistence length%
\begin{equation}
a=\frac{b_{1}}{kT}.
\end{equation}%
The simplest example of a filament under tension is that of a constant force
acting on one end and directed along the $x$-axis, with the second end
constrained to lie at the origin but otherwise free to rotate about it. The
energy of the filament thus picks up an additional contribution $-Fx$, where 
$x$ is the projection of the end to end vector on the direction of the
force. The problem is that $x$ is a complicated nonlinear function of all
the $\delta \omega $, and so the external force destroys the Gaussian
structure of the energy on which the generation of the $\delta \omega $
depends. One can simply incorporate the extra contribution to the energy by
treating it as a weighting factor, or bias, and suitably normalizing all
averages. Thus, for example, one can calculate the mean value of $x$ via 
\begin{equation}
\overline{x}=\frac{\sum xe^{fx}}{\sum e^{fx}}  \label{dumb-bias}
\end{equation}%
where the sum is over configurations generated via the Frenet algorithm
described above and $f\equiv F/kT$ is the force in units of inverse
length. For small forces, this works well. However, as the force increases,
the probability of generating highly extended configurations which are
heavily favored by the bias, is so small that the sampling becomes
ineffective. One therefore needs a way to bias the generation of
configurations towards those that are extended, so they can be sampled
appropriately.

\begin{figure}[ptb]
\includegraphics[width=5in]{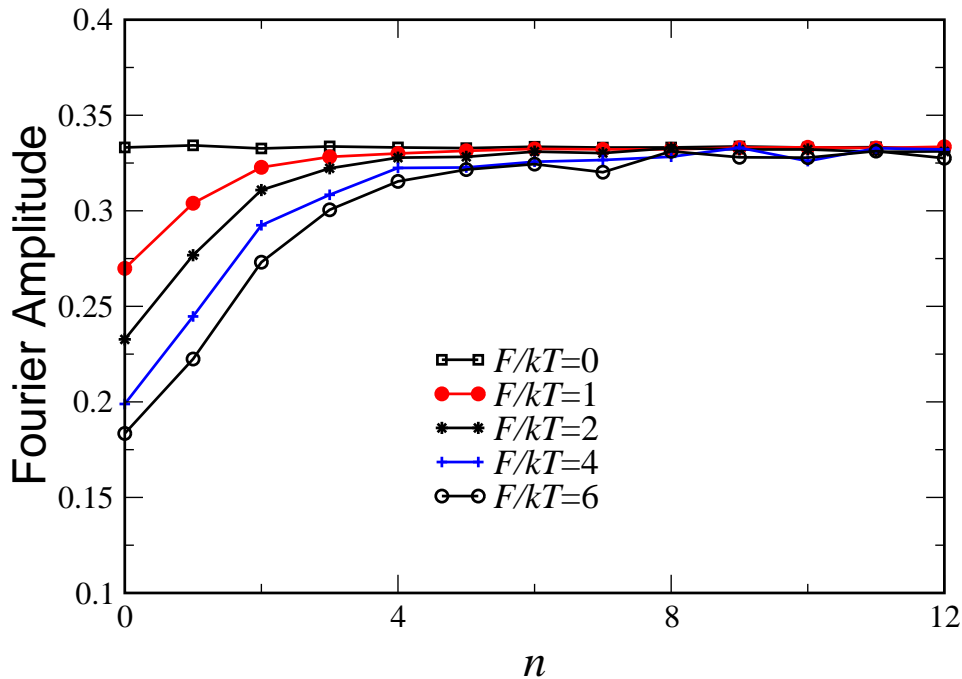} 
\caption{Amplitude of Fourier components of the curvature, as a function of
mode number $n$ for different forces $F/kT$, for $L=2\protect\pi$, $a=3$.
The lines are added to guide the eye.}
\label{kmoments}
\end{figure}

This task is not simple, since the connection between the $\delta \omega $'s
through which the filament is parameterized and the value of $x$                                     is very
complicated. We can gain insight into this by transforming to Fourier space
and looking at the expectation value of the moments of the $n$'th Fourier
component of the curvature, $\delta \widetilde{\omega }(n)$, using the above
algorithm, Eq. \ref{dumb-bias}, for small to intermediate forces where it is
still effective.  For an extensional force, we do not expect spontaneous
curvature to develop, and indeed the first moments of the Fourier components
all vanish. The second moments, however, do exhibit an interesting
structure, as seen in Fig. \ref{kmoments}, where the expectation value of
the squared amplitude of the low $n$ Fourier modes are presented for the
case $a=3$, $L=2\pi $. For $F=0$, the second moments are independent of $n$,
as expected. When the filament is subjected to a tensile force we see that
the small $n$ modes are suppressed, with the degree of suppression
increasing with $F$ and decreasing with $n$. In the case of large $a$, where
the fluctuations are in any case small, we observe that only the $n=0$ mode
is significantly affected by the force (not shown). The suppression of the
various curvature modes can be calculated analytically when fluctuations are
small, namely either in the large $FL$ limit, or for any $F$ provided $\chi
\equiv \sqrt{FL^{2}/(4akT)}\ll 1$ (see Appendix). The calculation
confirms the above-mentioned trends. We learn from this that the high
Fourier modes (or equivalently the small-scale fluctuations) of the filament
are essentially unaffected by the force, whereas the long-wavelength
fluctuations are significantly damped.

This suggests the possibility of biasing the Monte-Carlo by selectively
dampening the small-$n$ modes. The idea is to generate the $\delta 
\widetilde{\omega}$'s in Fourier space, with appropriate amplitudes for the
various modes, and construct the real-space $\delta\omega(s)$ via a Fast
Fourier Transform. Of course, this bias has to be corrected explicitly in
the weighting in the ensemble average. We do this in a self-consistent
fashion, choosing the amplitude of each Fourier mode to reproduce the
measured square magnitude of the mode. In practice, it is sufficient to do
this only for the first few modes, which are the ones most severely impacted
by the force. Also, in order to reduce the noise in the measurement of the
mode amplitudes, we only calculate the averages after every block of $N/10$
realizations and update accordingly, where $N$ is the total number of
realizations we intend to generate for these parameters. The force
dependence of the mode amplitudes is obtained by a stepwise increase of $F$,
where for each value of the force we use as a starting point the amplitudes
measured at the previous value.

There is one other degree of freedom of the problem that also must be
treated appropriately, namely the initial orientation of the filament at its
constrained end. This is because the force not only straightens the
filament, but also orients it toward its direction. We have employed two
different algorithms to deal with this problem. The simpler approach is to
simply bias the choice of initial orientation, and correct for this in the
weighting. The zero force distribution of the orientation angle, $\theta $,
of the tangent vector with respect to the direction of the force is $%
P(\theta )=sin\theta d\theta =d(\cos \theta )$. In the presence of the force
we modified this to $P(cos\theta )=exp(\alpha \cos \theta )d(\cos \theta ),$
where one can choose $\alpha $ to reproduce the measured mean of $cos\theta $%
.

A more sophisticated approach is to explicitly perform the integral over $%
\theta $. We generate a given configuration and calculate its end-to-end
distance $R$ and angle $\theta _{\text{e}nd}$ that the tangent to the end of
the filament makes with the direction of the end-to-end vector $\mathbf{R}$.
In order to evaluate the contribution of this realization to the average
elongation, Eq. \ref{dumb-bias}, one can perform a rigid rotation of this
configuration about the end point and calculate the contribution to the
statistical sum, from each angle of rotation $\theta $. The resulting
integrals over $\theta $ can be expressed in terms of $R$ and $\theta _{%
\text{e}nd}$ as 
\begin{equation}
\int_{0}^{\pi }\,d\theta sin\theta \{Rcos(\theta +\theta _{\text{e}%
nd}),1\}e^{fR\cos (\theta +\theta _{\text{e}nd})}
\end{equation}%
where the curly braces indicate the integrand for the numerator and
denominator of the ensemble sum, respectively. These integrals can easily be
evaluated numerically, using standard techniques. This second method is more
efficient, but more involved to implement.

\begin{figure}[ptb]
\includegraphics[width=3.2in]{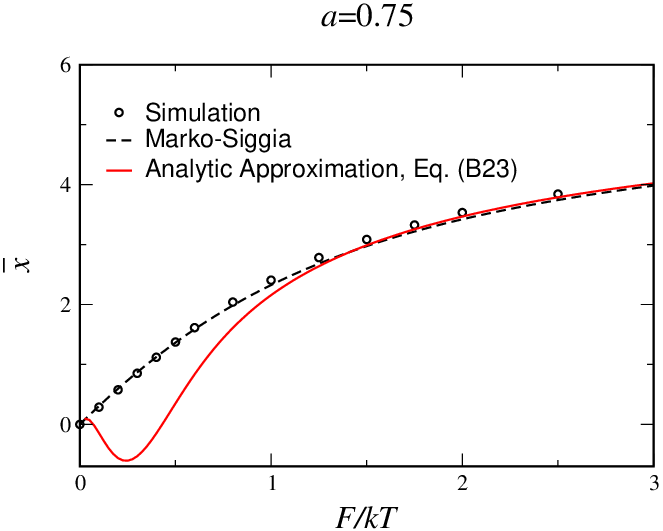} %
\includegraphics[width=3.2in]{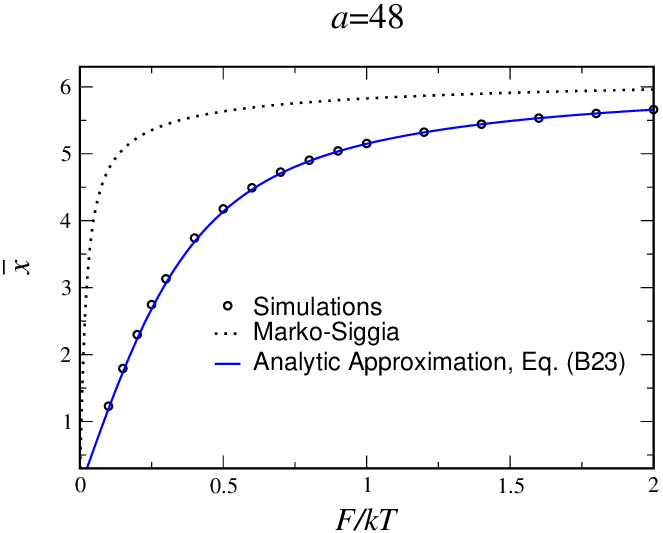} 
\caption{Extension along the force direction of a filament, as measured by
our Monte-Carlo procedure, together with the Marko-Siggia interpolation
formula and our analytic approximation, Eq. ({\ref{zunif}}) for $L=2\protect%
\pi$. Left: $a=0.75$; Right: $a=48.$}
\label{extenfig}
\end{figure}

Using these techniques, we have measured the force-extension curve $%
\overline{x}$ vs. $f$. We present the results for $a=.75$, $L=2\pi $ in the
left box of Fig. \ref{extenfig} together with the interpolation formula of
MS 
\begin{equation}
fa=\frac{\overline{x}}{L}+\frac{1}{4(1-\overline{x}/L)^{2}}-\frac{1}{4}
\label{MC}
\end{equation}%
We see that for this value of $a$ ($a\ll L)$ the agreement is very good. In
the right half of Fig. \ref{extenfig} we present the results for $a=48$,
again with $L=2\pi $. Here the MS formula does not work for small and
intermediate values of the force, since we are not in the long-chain limit ($%
L\gg a$) for which it was derived. On the other hand, for this case analysis
can also be done, as presented in the Appendix, and the analytic formula
derived there is in perfect agreement with the simulations (see the right
box of Fig. \ref{extenfig}). Notice that the MS expression systematically
overestimates the extension of the filament and approaches our results only
in the infinite force limit. The physical basis of the deviation is the
large orientational entropy which is not accounted for in the MS calculation.

Inspection of the right box of Fig. \ref{extenfig} suggests that both the
simulation and our analytical approximation (Eq. \ref{frey}) yield a finite
initial slope of the extension versus force curve in the rigid rod limit ($%
a\gg L$)$.$ Since the effective spring constant is proportional to the
inverse of the initial slope of this curve this suggests that the linear
response of the rigid rod to small stretching forces does not depend on the
persistence length $a$ and is a function of its length $L$ only. The above
expectation is confirmed by calculations reported in Appendix B where we
show that the corresponding spring constant is $3kT/L^{2}$. Obviously,
this behavior differs from the flexible chain limit ($a\ll L$) for which the
spring constant scales as\cite{PG} $kT/(aL).$ It also differs from the
predictions of Ref. [\onlinecite{frey2} ] where it was argued that the
effective spring constant diverges in the rigid rod limit, as $akT/L^{3}$%
. Notice that such a divergence is not expected for a single filament under
tension since there is a finite contribution to the susceptibility (and thus
to the response to the external force) due to orientational fluctuations,
even in the rigid rod limit. The difference between our result and that of
Ref. [\onlinecite{frey2} ] stems from the fact that the latter considers a
network in which filaments are not free to rotate and the only angular
averaging is over the randomly distributed orientations of different
filaments in the network.

Not only can we calculate the average $\overline{x}$, but we can also
measure the entire distribution $P(x)$. We show this in Fig. \ref%
{distribut-nowall}, for $a=0.75$ and $6$. In the former case, the
persistence length is much shorter than length of the filament and the
force-free distribution is nearly Gaussian. The peak of the distribution
moves to higher values of the extension and its width decreases with
increasing force. In the latter case (persistence length approximately equal
to the length of the filament), the force-free distribution is nearly flat
up to $x\simeq \pm 5,$ as expected for random sampling of orientations by a
nearly rigid rod (due to thermal fluctuations, the effective length of this
rod is somewhat shorter than its contour length, $2\pi $). The application
of small force (in the range $F/kT\leq 1$) orients the filament and
leads to dramatic change of the distribution. At larger forces, the
distribution changes slowly with the force, in a manner similar to the case $%
a=0.75$. \ The transition from orientation-dominated to stretching-dominated
behavior of nearly rigid rods is clearly observed in the right box of Fig. %
\ref{extenfig}, where pronounced deviations from the Marko-Siggia
formula are observed in the weak force regime (orientational effects are
not taken into account in the MS theory).

\begin{figure}[ptb]
\includegraphics[width=3.2in]{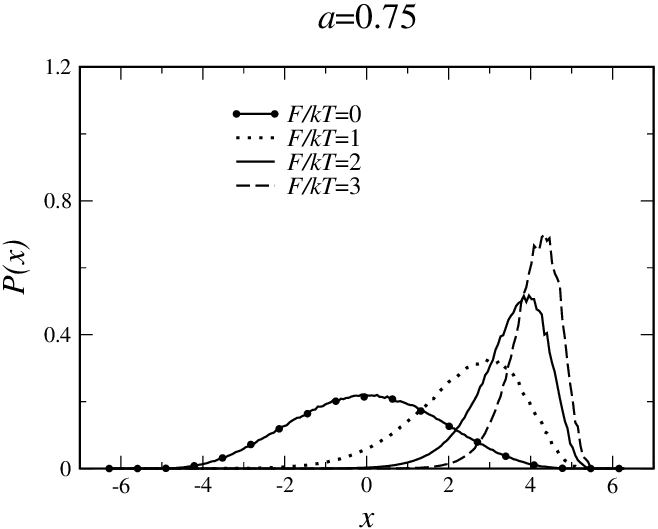} %
\includegraphics[width=3.2in]{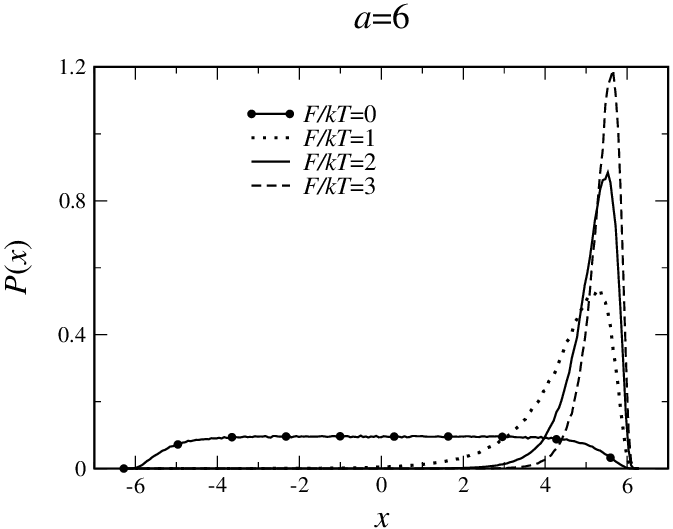} 
\caption{Distribution of the extension along the force direction, as
measured by our Monte-Carlo procedure, for the cases $f=0$, $1$, $2$ and $3$%
, with $L=2\protect\pi$. Left: $a=0.75$; Right: $a=6.$}
\label{distribut-nowall}
\end{figure}

\section{Steric Constraints}

\begin{figure}[ptb]
\includegraphics[width=3.2in]{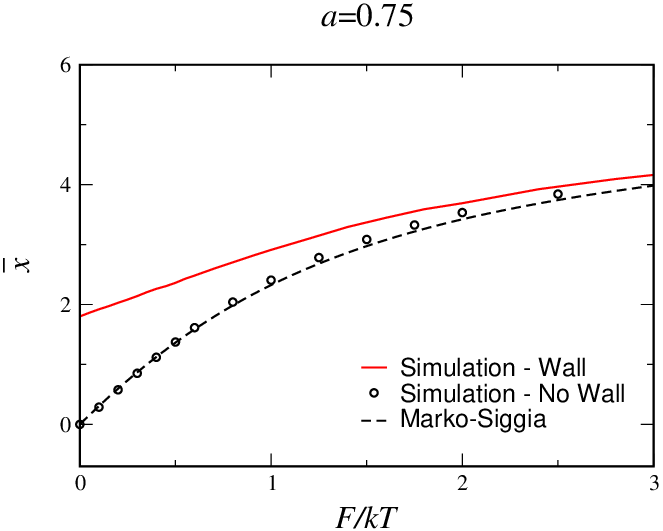} %
\includegraphics[width=3.2in]{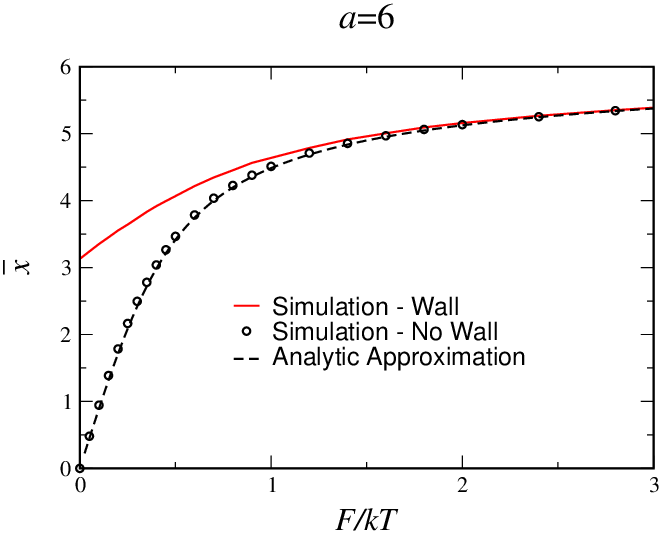} 
\caption{Force-extension curve with and without the presence of a wall
confining the filament to $x>0$ for $L=2\protect\pi$, measured via our
Monte-Carlo procedure. Left: $a=0.75$, together with the Marko-Siggia
formula; Right: $a=6$, together with our formula, Eq. ({\ref{zunif}}).}
\label{force-ext-wall}
\end{figure}

\begin{figure}[ptb]
\includegraphics[width=3.2in]{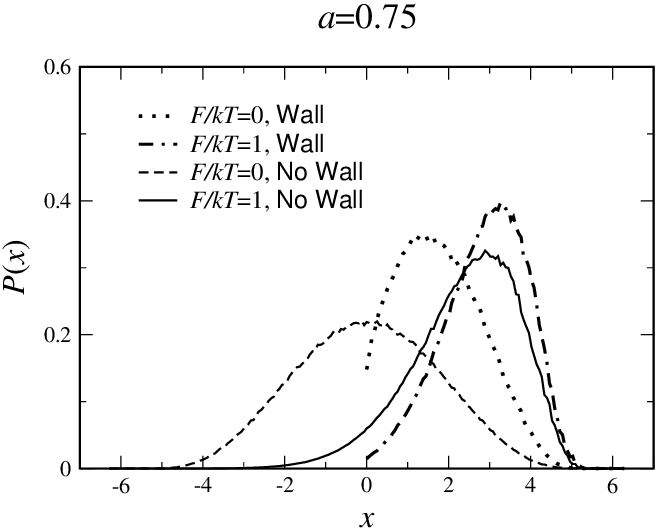} %
\includegraphics[width=3.2in]{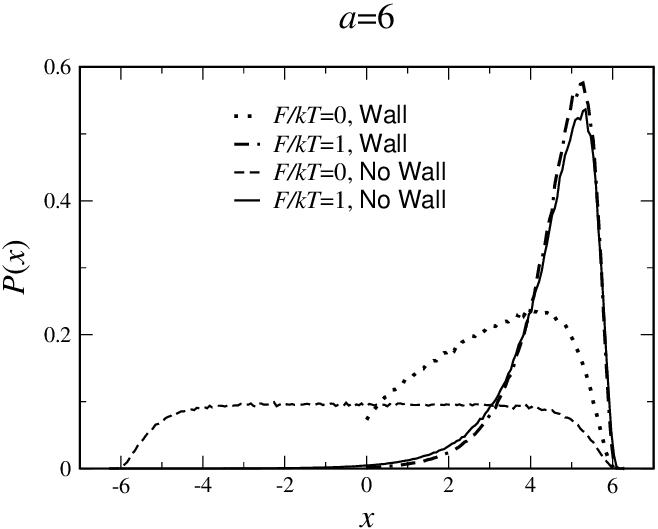} 
\caption{Distribution of the extension along the force direction with and
without the constraining wall, for $f=0$, $1$, with $L=2\protect\pi$. Left: $%
a=0.75$; Right: $a=6$. Distributions at higher forces are essentially
unaffected by the constraint.}
\label{xdist-wall}
\end{figure}

It is fairly straightforward to incorporate steric constraints into our
calculations. As an example, we consider the case where the filament is
confined to the upper half space $x>0$. After generating each configuration
according to the procedure outlined in the previous section, we simply
reject any configuration which violates the constraint. Our method of
integrating over the initial orientation angle $\theta $ can also be
employed. We have simply to compute the limits of the integral over $\theta $
which can be done straightforwardly given the configuration. We present the
results for the force-extension curve in Fig. \ref{force-ext-wall} and for
the distribution $P(x)$ in Fig. \ref{xdist-wall}. We see that, as one might
expect, the effect of the constraint is most pronounced at small $F$, where
the straightening/aligning of the curve is smallest. For vanishing $F$, the
average extension is positive as a result of the constraint. The constraint
basically carves out a ``hole'' in the distribution at small $x$, resulting
in a very asymmetric distribution for small force. For $F/kT>1$ the
distribution function is practically unaffected by the presence of the wall
and coincides with that shown in Fig. \ref{distribut-nowall}.

\section{Discussion}

In this work we used biased Monte-Carlo simulations, based on an extension of
our Frenet algorithm, to calculate the force-extension relations and the
distribution functions of the extension for filaments of arbitrary
persistence to contour length ratio. We showed that in the presence of
external force, curvature fluctuations are no longer Gaussian and that their
long wavelength Fourier modes are progressively suppressed by the force.
This effect was taken into account in a self-consistent manner in the
simulations by gradually increasing the force and adjusting the
corresponding Fourier amplitudes of curvature fluctuations. Analytic results
were obtained both for small fluctuations about the rigid rod limit, and in
the limit of large force. In the former case, we calculated the
force-induced renormalization of the length-scale-dependent bending moduli
that govern the various Fourier modes of curvature. As expected, the largest
effect is on the zeroth (constant curvature) mode; the corresponding modulus
increases quadratically and linearly with the force in the weak and strong
force limits, respectively. The underlying physical picture is reminiscent
of the suppression of excluded volume by tensile force in the self-avoiding
random walk model of polymers  - while large scale fluctuations are strongly
suppressed, small scale fluctuations (within the so called Pincus blobs\cite%
{pincus}) are not affected by the force.

For flexible filaments with $a\ll L$ our simulation reproduces the
Marko-Siggia force-extension curves. However, our simulations and analytic
results show that as the rigid rod limit $a\gg L$ is approached, the MS
expression increasingly overestimates the response of the filament at finite
values of the force, and approaches the correct result only asymptotically,
in the limit of infinite force. The discrepancy can be attributed to the
neglect of orientational effects (the constant curvature mode is not taken
into account) in the MS derivation which is strictly valid only in the $a/L$ 
$\rightarrow 0$ limit. These orientational effects are however seen in the
short-chain limit of the Gaussian model investigated recently by Winkler\cite%
{winkler}. The large force limit of the Gaussian model for short chains is
however very different from the worm-like chain model studied herein. For
long chains, the Gaussian model can reproduce either the weak or the strong
force limit, by appropriate choice of the persistence length, but not both.
This is an outcome of the length-scale dependence of the effective moduli
discussed above and in Appendix C.

Examination of the distribution functions for the extension shows that
increasing rigidity leads to a more rapid shift of the peak of the
distribution and narrowing of its width, with external force. The largest
deviations from Gaussian statistics of the distributions are observed for
stiff filaments in the weak to intermediate force range ($F/kT\leq 1$),
where fluctuations of initial orientation are not yet completely suppressed
and where wall effects play an important role. While it may be difficult to
approach the rigid rod limit in mechanical experiments on dsDNA that has a
persistence length of $500$ \AA , it can be readily accessed by experiments
on actin filaments\cite{albert} for which the persistence length of order $10
$ $\mu $m.

Finally, we would like to comment on possible extensions of the present
work. Since we only considered here filaments with symmetric cross section
and without spontaneous curvature, twist rigidity did not affect the
distribution of extension and played no role in our considerations. However,
since twist rigidity affects even the equilibrium distributions of ribbons
with strongly asymmetric cross sections\cite{jcp-kr}, it is expected that
both twist rigidity and spontaneous twist will affect the deformation of
such ribbons as well. Perhaps even more interesting is the effect of
spontaneous curvature. Indeed, we have shown that a helical filament with
sufficiently large twist to bending rigidity and radius to pitch ratios,
undergoes a sequence of stretching instabilities under the action of tensile
force, even in the absence of any thermal fluctuations\cite{prl-kr}. In the
presence of thermal fluctuations or mechanical noise these instabilities are
expected to lead to coexistence between different shapes of the deformed
object\cite{benedek}.

\begin{acknowledgments}
DAK and YR acknowledge the support of the Israel Science Foundation. We also
would like to thank J. Fineberg and G. Cohen for stimulating discussions.
\end{acknowledgments}

\appendix              

\section{Euler Angles}

Since we are interested in three dimensions, we work in terms of the Euler
angles $\theta$, $\phi$ and $\psi.$ For convenience we will orient the force
along the $x$-axis, i.e. along the direction $\theta=\pi/2,$ $\phi=0$. The
energy due to the force is then 
\begin{equation}
E_{F}=-Fx=-F\int_{0}^{L}dst_{3}^{x}=-F\int_{0}^{L}ds\sin(\theta(s))\cos
(\phi(s))
\end{equation}
The Euler angles satisfy the differential equations\cite{ring}: 
\begin{align}
\frac{d\theta}{ds} & =\omega_{1}\sin\psi+\omega_{2}\cos\psi  \notag \\
\frac{d\phi}{ds}\sin\theta & =-\omega_{1}\cos\psi+\omega_{2}\sin \psi  \notag
\\
\frac{d\psi}{ds}\sin\theta & =(\omega_{1}\cos\psi-\omega_{2}\sin\phi
)\cos\theta+\omega_{3}\sin\theta
\end{align}
We need to expand to quadratic order about the $s=0$ values $\theta_{0}$ and 
$\phi_{0}$, and to linear order about $\psi_{0}$, (since the energy does not
involve $\psi$ directly) giving 
\begin{align}
\frac{d\theta}{ds} & =\kappa-\lambda(\psi-\psi_{0}),  \notag \\
\frac{d\phi}{ds} & =\frac{\lambda+\kappa(\psi-\psi_{0})}{\sin\theta_{0}}-%
\frac{\cos\theta_{0}}{\sin^{2}\theta_{0}}\lambda(\theta-\theta _{0}),  \notag
\\
\frac{d\psi}{ds} & =-\frac{\cos\theta_{0}}{\sin\theta_{0}}\lambda+\omega_{3},
\end{align}
where $\kappa$ and $\lambda$ are the rotated curvatures 
\begin{align}
\kappa & \equiv\omega_{1}\sin\psi_{0}+\omega_{2}\cos\psi_{0},  \notag \\
\lambda & \equiv-\omega_{1}\cos\psi_{0}+\omega_{2}\sin\psi_{0},
\end{align}
Thus, to second order in the curvatures 
\begin{align}
\theta(s) & =\theta_{0}+\int_{0}^{s}ds^{\prime}\kappa(s^{\prime})+\frac{%
\cos\theta_{0}}{2\sin\theta_{0}}\left(
\int_{0}^{s}ds^{\prime}\lambda(s^{\prime})\right)
^{2}-\int_{0}^{s}\,ds^{\prime}\lambda(s^{\prime
})\int_{0}^{s^{\prime}}ds^{\prime\prime}\omega_{3}(s^{\prime\prime }), 
\notag \\
\phi(s) & =\phi_{0}+\frac{1}{\sin\theta_{0}}\int_{0}^{s}ds^{\prime}%
\lambda(s^{\prime})-\frac{\cos\theta_{0}}{\sin^{2}\theta_{0}}\left( \int
_{0}^{s}ds^{\prime}\kappa(s^{\prime})\right) \left( \int_{0}^{s}ds^{\prime
}\lambda(s^{\prime})\right)  \notag \\
& \ {}+\frac{1}{\sin\theta_{0}}\int_{0}^{s}\,ds^{\prime}\kappa(s^{\prime
})\int_{0}^{s^{\prime}}ds^{\prime\prime}\omega_{3}(s^{\prime\prime}).
\end{align}

\section{Force-Extension Curve}

We will first compute the force-extension curve, via the partition function.
We do this in two stages, depending on the value of the parameter $\chi
\equiv\sqrt{FL^{2}/4a}$.

\subsection{ Small $\protect\chi$ ($\protect\chi\ll1,$ arbitrary $fL)$}

Here we write 
\begin{equation}
E_{F}=-FL\sin \theta _{0}\cos \phi _{0}+F\delta 
\end{equation}%
where $\delta $ is given by 
\begin{align}
\delta & =-\int_{0}^{L}\,ds\left[ \cos \theta _{0}\cos \phi
_{0}\int_{0}^{s}\kappa -\sin \phi _{0}\int_{0}^{s}\lambda -\frac{1}{2}\sin
\theta _{0}\cos \phi _{0}\left[ \left( \int_{0}^{s}\kappa \right)
^{2}+\left( \int_{0}^{s}\lambda \right) ^{2}\right] \right.   \notag \\
& \ \hspace{0.5in}\left. -\cos \theta _{0}\cos \phi
_{0}\int_{0}^{s}\,ds^{\prime }\lambda (s^{\prime })\int_{0}^{s^{\prime
}}\omega _{3}-\sin \phi _{0}\int_{0}^{s}\,ds^{\prime }\kappa (s^{\prime
})\int_{0}^{s^{\prime }}\omega _{3}\right] 
\end{align}%
We expand $e^{-F\delta /kT}$ to quadratic order, yielding 
\begin{align}
Z& =\int d\Omega _{0}\mathcal{D}\kappa \mathcal{D}\lambda \exp {\left[ -%
\frac{a}{2}\int_{0}^{L}(\kappa ^{2}+\lambda ^{2})+fL\sin \theta _{0}\cos
\phi _{0}\right] }  \notag \\
& \ \times \left[ 1-f\delta +\frac{f^{2}}{2}\delta ^{2}\right] 
\end{align}%
where $\int d\Omega _{0}\equiv \frac{1}{4\pi }\int_{0}^{\pi }\sin \theta
_{0}d\theta _{0}\int_{0}^{2\pi }d\phi _{0}$. The functional integrals are
all Gaussian and can be easily done by going to Fourier space. For example, 
\begin{equation}
\kappa (s)=\kappa _{0}+\sum_{n=1}^{\infty }\kappa _{n}^{c}\cos (\frac{2\pi ns%
}{L})+\sum_{n=1}^{\infty }\kappa _{n}^{s}\sin (\frac{2\pi ns}{L})
\end{equation}%
so that 
\begin{equation}
\int_{0}^{L}\left( \int_{0}^{s}ds^{\prime }\kappa \right) ^{2}ds=\frac{L^{3}%
}{3}\kappa _{0}^{2}+L^{3}\sum_{n=1}^{\infty }\frac{(\kappa
_{n}^{c})^{2}+3(\kappa _{n}^{s})^{2}}{2(2\pi n)^{2}}+mixed\,\,\,terms
\end{equation}%
where the mixed terms will vanish latter upon functional integration and are
omitted in the following. Defining 
\begin{equation}
\mathcal{D}_{0}\kappa \equiv \mathcal{D}\kappa e^{-\frac{a}{2}%
\int_{0}^{L}\kappa ^{2}ds}=\mathcal{D}\kappa e^{-\frac{aL}{2}\kappa _{0}^{2}-%
\frac{aL}{4}\sum_{n=1}^{\infty }\left( (\kappa _{n}^{c})^{2}+(\kappa
_{n}^{s})^{2}\right) }
\end{equation}%
and carrying out the functional integration by integrating over all the
Fourier components 
\begin{equation}
\mathcal{D}\kappa \equiv d\kappa _{0}\prod_{i=1}^{\infty }d\kappa
_{i}^{c}\prod_{j=1}^{\infty }d\kappa _{j}^{s},
\end{equation}%
we get 
\begin{equation}
\int \mathcal{D}_{0}\kappa \int_{0}^{L}\left( \int_{0}^{s}ds^{\prime }\kappa
\right) ^{2}ds=\mathcal{N}\frac{L^{2}}{2a}
\end{equation}%
where $\mathcal{N}=\int \mathcal{D}_{0}\kappa \,$ is the (normalization)
integral over the exponential factor only. Similarly, 
\begin{equation*}
\int \mathcal{D}_{0}\kappa \left[ \int_{0}^{L}\left( \int_{0}^{s}\kappa
\right) ds\right] ^{2}=\int \mathcal{D}_{0}\kappa \left[ \frac{\kappa
_{0}L^{2}}{2}+\sum_{n=1}^{\infty }\frac{\kappa _{n}^{s}L^{2}}{2\pi n}\right]
^{2}=\mathcal{N}\frac{L^{3}}{3a}.
\end{equation*}%
The functional integration with respect to $\lambda $ is carried out in the
same manner. Putting this all together, we get 
\begin{equation}
Z=\mathcal{N}^{2}\int d\Omega _{0}e^{fL\sin \theta _{0}\cos \phi _{0}}\left[
1-\frac{fL^{2}}{2a}\sin \theta _{0}\cos \phi _{0}+\frac{f^{2}L^{3}}{6a}%
\left( \cos ^{2}\theta _{0}\cos ^{2}\phi _{0}+\sin ^{2}\phi _{0}\right) %
\right] 
\end{equation}%
The integrals over solid angle are as follows: 
\begin{align}
\int d\Omega _{0}e^{fL\sin \theta _{0}\cos \phi _{0}}& =\frac{\sinh fL}{fL} 
\notag \\
\int d\Omega _{0}e^{fL\sin \theta _{0}\cos \phi _{0}}\sin \theta _{0}\cos
\phi _{0}& =\frac{\cosh fL}{fL}-\frac{\sinh fL}{(fL)^{2}}  \notag \\
\int d\Omega _{0}e^{fL\sin \theta _{0}\cos \phi _{0}}\cos ^{2}\theta
_{0}\cos ^{2}\phi _{0}& =\frac{1+\cosh fL}{(fL)^{2}}-\frac{2\sinh fL}{%
(fL)^{3}}  \notag \\
\int d\Omega _{0}e^{fL\sin \theta _{0}\cos \phi _{0}}\sin ^{2}\phi _{0}& =%
\frac{\cosh fL-1}{(fL)^{2}}
\end{align}%
Collecting all terms we finally get 
\begin{equation}
Z=\mathcal{N}^{2}\left\{ \frac{\sinh fL}{fL}-\frac{L}{6a}\left[ \cosh fL-%
\frac{\sinh fL}{fL}\right] \right\}   \label{z-smallchi}
\end{equation}

From this, we can differentiate to get the average projection of the
end-to-end vector on the $x$-axis, $\overline{x}=d(\ln Z)/df$ 
\begin{equation}
\overline{x}=\frac{L}{\tanh fL}-\frac{1}{f}-\frac{2\chi ^{2}}{3f}\left[ 
\frac{1}{\tanh fL}-\frac{fL}{\sinh ^{2}fL}\right]   \label{frey}
\end{equation}%
At large $fL$ (but still small $\chi $), this goes as $L-L^{2}/(6a)-1/f$.
For small $fL$ this projection goes linearly with $f$ as $\overline{x}%
\approx fL^{2}/3(1-L/(3a))$ and we conclude that in the limit $a\gg L$ the
effective ``spring constant'' is $3kT/L^{2}$.

\subsection{Large $fL$ (arbitrary $\protect\chi$)}

We start by expanding $\theta_{0}$, $\phi_{0}$ about the values $\theta
_{0}^{\ast}$, $\phi_{0}^{\ast}$ which minimize $E_{F}$. To lowest order, 
\begin{align}
\theta_{0}^{\ast} & =\frac{\pi}{2}-\frac{1}{L}\int_{0}^{L}ds\int_{0}^{s}ds^{%
\prime}\kappa  \notag \\
\phi_{0}^{\ast} & =-\frac{1}{L}\int_{0}^{L}ds\int_{0}^{s}ds^{\prime}\lambda
\end{align}
so that the expansion reads 
\begin{align}
E_{F} & =-FL-\frac{F}{2L}\left[ \left(
\int_{0}^{L}ds\int_{0}^{s}ds^{\prime}\kappa\right) ^{2}+\left(
\int_{0}^{L}ds\int_{0}^{s}ds^{\prime}\lambda\right) ^{2}\right] +\frac{FL}{2}%
\left[ \left( \theta_{0}-\theta_{0}^{\ast}\right) ^{2}+\left(
\phi_{0}-\phi_{0}^{\ast }\right) ^{2}\right]  \notag \\
& \ +\frac{F}{2}\int_{0}^{L}\,ds\left[ \left(
\int_{0}^{s}ds^{\prime}\kappa\right) ^{2}+\left(
\int_{0}^{s}ds^{\prime}\lambda\right) ^{2}\right]
\end{align}
Let us now consider the functional integral over $\kappa$, the integral over 
$\lambda$ being exactly the same. We have 
\begin{align}
Z_{\kappa} & =\int\mathcal{D}\kappa e^{-\frac{a}{2}\int_{0}^{L}ds\kappa ^{2}-%
\frac{f}{2}\int_{0}^{L}\,ds\left( \int_{0}^{s}ds^{\prime}\kappa\right) ^{2}+%
\frac{f}{2L}\left[ \int_{0}^{L}\,ds\int_{0}^{s}ds^{\prime}\kappa\right] ^{2}}
\notag \\
& =\int\mathcal{D}\kappa\exp\left\{ -\frac{aL}{2}\kappa_{0}^{2}-\frac{aL}{4}%
\sum_{n=1}^{\infty}\left( (\kappa_{n}^{c})^{2}+(\kappa_{n}^{s})^{2}\right)
\right.  \notag \\
& \ \left. {}-\frac{fL^{3}}{2}\left[ \frac{\kappa_{0}^{2}}{12}+\sum
_{n=1}^{\infty}\left( \frac{(\kappa_{n}^{c})^{2}}{2(2\pi n)^{2}}+\frac{%
(\kappa_{n}^{s})^{2}}{2(2\pi n)^{2}}-\frac{2\kappa_{0}\kappa_{n}^{c}}{(2\pi
n)^{2}}\right) \right] \right\}
\end{align}
We can diagonalize this expression by shifting $\kappa_{n}^{c}$, defining 
\begin{equation}
C_{n}\equiv\kappa_{n}^{c}-\frac{b_{n}}{a_{n}}\kappa_{0}
\end{equation}
with 
\begin{equation}
a_{n}\equiv\frac{aL+b_{n}}{2};\quad\quad b_{n}\equiv\frac{fL^{3}}{(2\pi
n)^{2}}
\end{equation}
yielding an effective modulus of the zero mode 
\begin{equation}
a_{0}=a+\frac{fL^{2}}{12}-\sum_{n=1}^{\infty}\frac{b_{n}^{2}}{a_{n}}=\frac{%
a\chi}{\tanh\chi}   \label{a0highf}
\end{equation}
We can now calculate $Z_{\kappa}$: 
\begin{align}
Z_{\kappa}/Z_{\kappa,0} & =(a_{0}/a)^{-1/2}\prod_{n=1}^{\infty}\frac{aL}{%
2a_{n}}  \notag \\
& =(a_{0}/a)^{-1/2}\prod_{n=1}^{\infty}\left( 1+\frac{\chi^{2}}{(\pi n)^{2}}%
\right) ^{-1}  \notag \\
& =(a_{0}/a)^{-1/2}\frac{\chi}{\sinh\chi}  \notag \\
& =\left[ \frac{2\chi}{\sinh2\chi}\right] ^{1/2}
\end{align}
where $Z_{\kappa,0}$ is the $\kappa$ partition function at $f=0$. The entire
partition function is then 
\begin{equation}
Z=\mathcal{N}^{2}\frac{e^{fL}}{2fL}\frac{2\chi}{\sinh2\chi}
\end{equation}
For small $\chi$, this yields 
\begin{equation}
Z\approx\mathcal{N}^{2}\frac{e^{fL}}{2fL}\left( 1-\frac{2\chi^{2}}{3}\right)
,
\end{equation}
which agrees precisely with the large $f$ expansion of Eq. (\ref{z-smallchi}%
).

The average extension in this second region ($fL\gg1$) is given by 
\begin{equation}
\overline{x}=L-\frac{1}{2f}-\frac{\chi}{f\tanh2\chi}
\end{equation}
which for small $\chi$ gives $\overline{x}\approx L-1/f-L^{2}/(6a),$ in
accord with the large $fL$ limit of the first region ($\chi\ll1$). At large $%
fL$, it agrees with the Marko--Siggia result, Eq. (\ref{MC}). A uniform
approximation for the force-extension relation is then 
\begin{equation}
\overline{x}\approx-\frac{1}{2f}-\frac{\chi}{f\tanh2\chi}+\frac{L}{\tanh fL}-%
\frac{2\chi^{2}}{3f}\left( \frac{1}{\tanh fL}-\frac{fL}{\sinh^{2}fL}%
-1\right) .   \label{zunif}
\end{equation}
This result is confirmed by simulations, as shown in data for $a=48$, Fig. %
\ref{extenfig}. It even works well when the persistence length slightly
shorter than the chain length, as can be seen in the data for $a=6$ in the
right half of Fig. \ref{force-ext-wall}. Since our analysis is limited to
the small fluctuation (stiff rod and/or large force) regime, we expect it to
break down for very short persistence lengths in the small force limit. This
is clearly observed in the case of $a=0.75$, shown in the left half of Fig. %
\ref{extenfig}.

\section{Moduli}

\begin{figure}[ptb]
\includegraphics[width=4.0in]{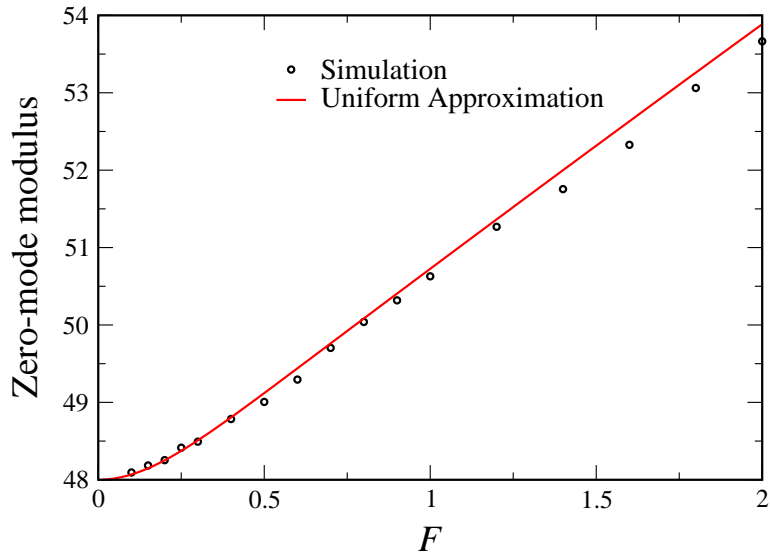} 
\caption{Comparison of simulations for the zero-mode modulus as a function
of the force $F/kT$ to our uniform approximation, Eq. ({\ref{a0unif}}).}
\label{0modeamp}
\end{figure}

We now turn to the computation of the effective moduli of the various modes.
We first treat the zero modes, starting as above with the small $\chi$
limit. We proceed by calculating a modified partition function, $\tilde{Z}$
with a modified modulus of the zero modes, $\tilde{a}$, replacing the bare
modulus $a$. Then 
\begin{equation}
\frac{L}{2}\langle\kappa_{0}^{2}+\lambda_{0}^{2}\rangle=\left. -\frac{d\ln 
\tilde{Z}}{d\tilde{a}}\right| _{\tilde{a}=a}
\end{equation}
Tracking the changes induced in the partition function calculation above, we
find 
\begin{equation}
\tilde{Z}=\frac{a}{\tilde{a}}\mathcal{N}^{2}\left\{ \frac{\sinh fL}{fL}%
-fL^{2}\left[ \left( \frac{1}{3\tilde{a}}+\frac{1}{6a}\right) -\left( \frac{1%
}{4\tilde{a}}+\frac{1}{12a}\right) \right] \left[ \frac{\cosh fL}{fL}-\frac{%
\sinh fL}{(fL)^{2}}\right] \right\}
\end{equation}
giving 
\begin{equation}
\ln\tilde{Z}=\ln a-\ln\tilde{a}+2\ln\mathcal{N}+\ln\frac{\sinh fL}{fL}%
-fL^{2}\left( \frac{1}{12\tilde{a}}+\frac{1}{24a}\right) \left( \frac{\cosh
fL}{\sinh fL}-\frac{1}{fL}\right)
\end{equation}
so that 
\begin{equation}
\left. \frac{d\ln\tilde{Z}}{d\tilde{a}}\right| _{a}=-\frac{1}{a}+\frac{fL^{2}%
}{12a^{2}}\left[ \frac{\cosh fL}{\sinh fL}-\frac{1}{fL}\right]
\end{equation}
This implies an effective zero mode modulus of 
\begin{equation}
a_{0}=a+\frac{fL^{2}}{12a}\left( \frac{1}{\tanh fL}-\frac{1}{fL}\right)
\end{equation}
At small and large forces, this reads 
\begin{equation}
a_{0}\approx\left\{ 
\begin{array}{cl}
a+f^{2}L^{3}/(36a) & fL\ll1 \\ 
a+fL^{2}/(12a) & fL\gg1%
\end{array}
\right.   \label{a0lims}
\end{equation}
At large $f$, the effective zero mode modulus was calculated above, Eq. (\ref%
{a0highf}). The small $\chi$ limit of this agrees with the large force limit
of Eq. (\ref{a0lims}), as it must. A uniform approximation for $a_{0}$ reads 
\begin{equation}
a_{0}=\frac{fL^{2}}{12a}\left( \frac{e^{-fL}}{\sinh fL}-\frac{1}{fL}\right)
+a\frac{\chi}{\tanh\chi}   \label{a0unif}
\end{equation}
This result is confirmed by the simulations as seen in Fig. \ref{0modeamp}.

Now we turn to the finite wave-number modes. The interesting thing to note
here is that the \textit{cos} modes renormalize differently than the \textit{%
sin} modes, due to the coupling of the zero mode to the \textit{cos} modes.
At large force, the effective modulus of the \textit{sin} modes is the
trivial 
\begin{equation}
a_{n}^{s} = a + \frac{fL^{2}}{(2\pi n)^{2}}
\end{equation}
so that the long wavelength modes are the most effected by the force. The
effective modulus of the \textit{cos} modes follows from 
\begin{equation}
\langle(\kappa_{n}^{c})^{2} \rangle= \langle C_{n}^{2} \rangle+ \frac{%
b_{n}^{2}}{a_{n}^{2}} \langle\kappa_{0}^{2} \rangle= \frac{1}{a_{n}} + \frac{%
b_{n}^{2}}{a_{n}^{2}}\frac{1}{a_{\textit{\tiny eff},0}L}
\end{equation}
with 
\begin{equation}
a^{c}_{n}=\frac{2}{L\langle(\kappa_{n}^{c})^{2} \rangle}
\end{equation}
This expression is fairly messy, so let us examine its various limits. For
large enough mode number, the effective modulus is $a + fL^{2}/(2\pi n)^{2}$
as for the \textit{sin} modes. For intermediate forces, all modes are large
enough for this to be the case. For strong forces, however, the effective
modulus is $L\sqrt{fa}/4$ up to some mode number, at which point it crosses
over to the previous result.


\begin{thebibliography}{99}
\bibitem{bustamante} C. Bustamante, Z. Bryant and S. B. Smith, \textit{Nature%
} \textbf{421}, 423 (2003).

\bibitem{albert} A. Ott, M. Magnasco, A. Simon and A. Libchaber, Phys. Rev.
E 48, R1642 (1993).

\bibitem{fred} F. C. MacKintosh, J. K\"{a}s and P. A. Janmey, \prl\textbf{75}%
, 4425 (1995).

\bibitem{sackmann} B. Hinner, M. Tempel, E. Sackmann, K. Kroy and E. Frey, %
\prl  \textbf{81}, 2614 (1998).

\bibitem{ms1} C. Bustamante, J. F. Marko, E. D. Siggia and S. Smith, \textit{%
Science} 265, 1599 (1994).

\bibitem{frey2} K. Kroy and E. Frey, \prl  \textbf{77}, 306 (1996).

\bibitem{frey1} J. Wilhelm and E. Frey, \prl  \textbf{77}, 2581 (1996).

\bibitem{sinha} J. Samuel and S. Sinha, \pre \textbf{66}, 050801 (2002).

\bibitem{dhar} A. Dhar and D. Chaudhauri, \prl \textbf{89}, 065502 (2002).

\bibitem{jcp-kr} D. A. Kessler and Y. Rabin, \textit{J. Chem. Phys.} \textbf{%
118}, 897 (2003).

\bibitem{winkler} R. G. Winkler, \textit{J. Chem. Phys.} \textbf{118}, 2919
(2003).

\bibitem{thesis} J. F. Allemand, Ph.D Thesis (Universit\'{e} Pierre et Marie
Curie, 1997).

\bibitem{Kats} Y. Kats, D. A. Kessler and Y. Rabin, \pre\textbf{65},
020801(R) (2002).

\bibitem{MS94} J. F. Marko and E. D. Siggia, \textit{Macromolecules} \textbf{%
27}, 981 (1994).

\bibitem{odijk} T. Odijk, \textit{Macromolecules} \textbf{28}, 7016 (1995).

\bibitem{ha} B. Y. Ha and D. Thirumalai, \textit{J. Chem. Phys.} \textbf{106}%
, 4243 (1997).

\bibitem{Helixpre} S. V. Panyukov and Y. Rabin, \prl\textbf{85}, 2404
(2000); \pre\textbf{62}, 7135 (2000).

\bibitem{Love} A. E. H. Love, \textit{A Treatise on the Mathematical Theory
of Elasticity} (Dover, New York, 1944).

\bibitem{PG} P. G. deGennes, \textit{Scaling Concepts in Polymer Physics}
(Cornell University Press, Ithaca, 1979).

\bibitem{pincus} P. Pincus, \textit{Macromolecules} \textbf{10}, 210 (1977).

\bibitem{ring} S. V. Panyukov and Y. Rabin, \pre\textbf{64}, 011911 (2001).

\bibitem{prl-kr} D. A. Kessler and Y. Rabin, \textit{Phys. Rev. Lett.} 
\textbf{90}, 024301 (2003).

\bibitem{benedek} B. Smith, Y. V. Zastavker and G. B. Benedek, \textit{Phys.
Rev. Lett.} \textbf{87}, 278101 (2001).
\end{thebibliography}
\end{document}